\documentclass[fleqn,10pt]{wlscirep}
\usepackage[utf8]{inputenc}
\usepackage[T1]{fontenc}
\title{Spin Hall Nano-Antenna}

\author[1]{Raisa Fabiha}
\author[2]{Pratap Kumar Pal}
\author[1]{Michael Suche}
\author[2]{Amrit Kumar Mondal}
\author[1]{Erdem Topsakal}
\author[2]{Anjan Barman}
\author[1,$\dag$]{Supriyo Bandyopadhyay}
\affil[1]{Department of Electrical and Computer Engineering, Virginia Commonwealth University, Richmond, VA 23284, USA}
\affil[2]{Department of Condensed Matter and Materials Physics, S. N. Bose National Centre for Basic Sciences, Block–JD, Sector–III, Salt Lake, Kolkata, 700106, India}

\affil[$\dag$]{sbandy@vcu.edu}

\begin{abstract}
{\bf The spin Hall effect \cite{dyakonov,hirsch} is a celebrated phenomenon in spintronics and magnetism that has found numerous applications in digital electronics (memory and logic), but very few in analog electronics. Practically, the only analog application in widespread use is the spin Hall nano-oscillator (SHNO) \cite{chen} that delivers a high frequency alternating current or voltage to a load.  Here, we report its analogue -- a spin Hall nano-antenna (SHNA) that radiates a high frequency electromagnetic wave (alternating electric/magnetic fields) into the surrounding medium. It can also radiate an acoustic wave in an underlying substrate if the nanomagnets are made of a magnetostrictive material. That makes it a {\it dual} electromagnetic/acoustic antenna. The SHNA is made of an array of ledged magnetostrictive nanomagnets deposited on a substrate, with a heavy metal nanostrip underlying/overlying the ledges. An alternating charge current passed through the nanostrip generates an alternating spin-orbit torque in the nanomagnets via the spin Hall effect  which makes their magnetizations oscillate in time with the frequency of the current, producing confined spin waves (magnons), which radiate electromagnetic waves (photons) in space with the same frequency as the ac current. Despite being much smaller than the radiated wavelength,  the SHNA surprisingly does not act as a point source which would radiate isotropically. Instead, there is clear directionality (anisotropy) in the radiation pattern, which is frequency-dependent. This is due to the (frequency-dependent) intrinsic anisotropy in the confined spin wave patterns generated within the nanomagnets, which effectively endows the ``point source'' with internal anisotropy.}
\end{abstract}

\begin{document}

\flushbottom
\maketitle
%
%
\thispagestyle{empty}

\section{Introduction}
The spin Hall effect is a remarkable phenomenon in spintronics that is used to inject spin currents into  ferromagnets by passing charge currents through an underlying heavy metal or topological insulator \cite{liu,mellnik}. This results in a spin-orbit torque within the ferromagnet which is harnessed to switch its magnetization, thereby making it an efficient mechanism to write bits into non-volatile magnetic memory \cite{memory}. The same effect has also been leveraged to implement combinational Boolean logic using magnetic switches \cite{logic}. These are digital applications. Analog applications of the spin Hall effect are few and far between, with the most notable application being in the spin Hall nano-oscillator (SHNO) which is used in microwave assisted magnetic recording \cite{MAMR} and neuromorphic computing \cite{chen}, among others.

\begin{figure}[!h]
\centering
\includegraphics[width=0.99\textwidth]{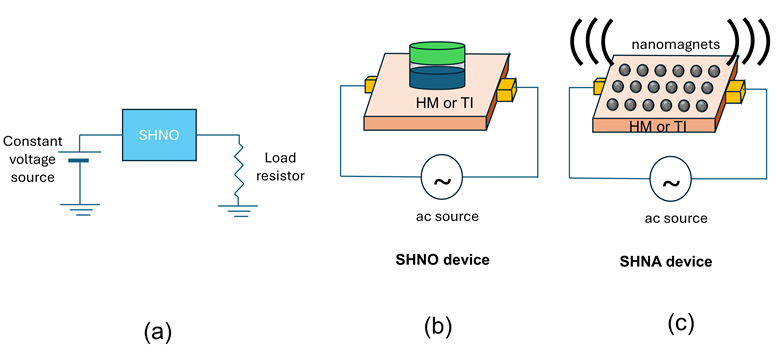}
\caption{(a) A spin Hall nano-oscillator (SHNO) acts as a time-varying resistor that delivers a time-varying current to a load resistor connected in series with it when both are powered by a dc voltage source. (b) Schematic of a SHNO device actuated by passing an alternating current through a heavy metal (HM) or topological insulator (TI) and no external magnetic field present. (c) Schematic of a spin Hall nano antenna (SHNA) where, again, no external magnetic field is needed.}
\label{fig:SHNO}
\end{figure}

SHNOs are high frequency oscillators that deliver a high frequency current to a load resistor \cite{chen} as shown in Fig. \ref{fig:SHNO}(a). The SHNO device usually consists of a magnetic tunnel junction (MTJ) whose soft layer is placed in contact with a heavy metal layer (HM) or a topological insulator (TI). An external magnetic field sets up magnetization precession in the soft layer to produce spin waves, and the precession is sustained by passing a direct current through the HM or TI, which generates an anti-damping spin-orbit torque on the soft layer of the MTJ via the spin Hall effect. This torque sustains the oscillations in time. Because the magnetization of the soft layer oscillates in time, the MTJ acts as an oscillating resistor that will deliver a time varying current to a load resistor when both are placed in series with a constant voltage source [see Fig. \ref{fig:SHNO}(a)].

The disadvantage of this approach is the need for an external magnetic field to set up the magnetization precession or spin waves. Field-free oscillators, working on a somewhat different principle, have been proposed \cite{abeed}, but not yet experimentally demonstrated. We can also drive an alternating (instead of direct) current through the HM or TI [see Fig. \ref{fig:SHNO}(b)] to generate an {\it alternating} (instead of time-invariant) spin-orbit torque in the soft layer. That will induce either back-and-forth domain wall motion \cite{fert} or magnetization precession to generate spin waves (magnetization oscillations) in the soft layer, predominantly at the frequency of the ac current, and without the need for any magnetic field. The MTJ will then once again act as an oscillating resistor that will deliver an alternating current to a load with the frequency of the injected current. This is, of course, not a true SHNO since we are not converting dc power from a dc current/voltage to ac power, which an ``oscillator'' should, but it will nonetheless produce the oscillating resistor -- this time without the need for an external magnetic field. The magnetization oscillation generated by an alternating spin-orbit torque may have other uses as well, but here we have leveraged it to implement an ``antenna'' that radiates electromagnetic waves into the surrounding medium with the frequency of the injected alternating current. The fact that an alternating spin-orbit torque generated by passing an alternating current through a HM layer produces magnetization oscillation of the same frequency in a nanomagnet in contact with the HM, has been verified experimentally by us in the past \cite{ahsanul}.

\begin{figure}[!h]
\centering
\includegraphics[width=0.99\textwidth]{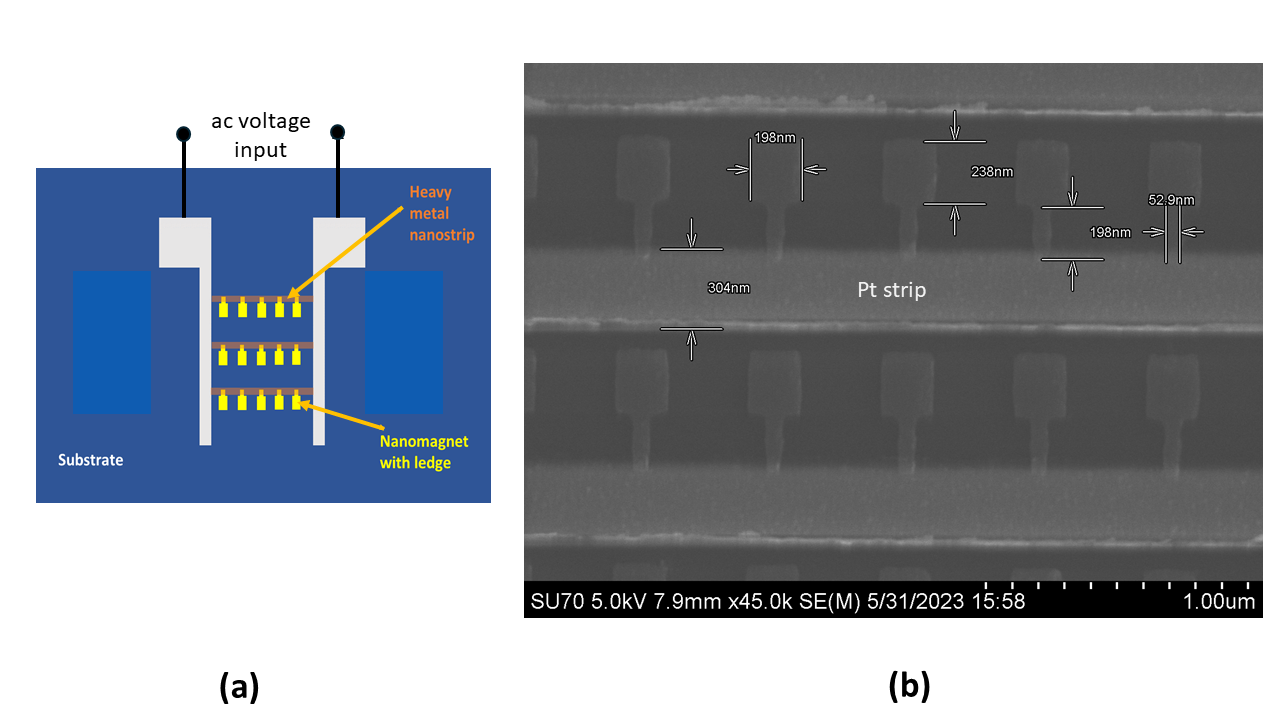}
\caption{(a) Schematic of the spin Hall nano-antenna device. (b) Scanning electron micrograph of the device showing the various structural dimensions).}
\label{fig:SHNA}
\end{figure}

An {\it antenna} is very distinct from an {\it oscillator}.  While the former  delivers an ac current or voltage to a load resistor, the latter ``radiates'' an ac signal (electromagnetic wave) into the surrounding medium, as shown in Fig. \ref{fig:SHNO}(c). Various types of modulation schemes (frequency modulation, phase-shift keying, frequency-shift keying, etc.) of spin torque (and by extension, spin Hall) nano-oscillators have been demonstrated for signal processing \cite{litvinenko,manfrini1,manfrini2,pufall,muduli,pogoryelov}, but there has been no demonstration of antenna functionality (wireless signal transmission over long distances). Here, we report  a {\bf spin Hall nano-antenna (SHNA)} made of nanomagnets placed in contact with a HM nanostrip through which an ac current is made to flow. The resulting  alternating spin-orbit torque sets up magnetization oscillations (confined spin waves) in the nanomagnets at the frequency of the current. These spin waves radiate an electromagnetic wave  into the surrounding medium at the  frequency of the ac current, thereby implementing an antenna. It is an extreme sub-wavelength antenna whose dimensions (sub-mm) are much smaller than the radiated wavelength at the frequencies we have tested, which are between 1 and 10 GHz (free space wavelength 3-30 cm), and yet it radiates efficiently. Had it been a conventional antenna, its radiation efficiency would have been very poor. This is an entirely new application of spin-orbit torque and spin Hall effect, and it can spawn a new genre of antennas.

\section{Experiment}

\subsection{Sample description}

The schematic of our SHNA device is shown in Fig. \ref{fig:SHNA}(a). It consists of linear periodic arrays of ``ledged'' rectangular nanomagnets deposited on a LiNbO$_3$ substrate, with a Pt nanostrip underlying the ledges as shown in Fig. \ref{fig:SHNA}.  The Pt nanostrip is $\sim$300 nm wide and 5 nm thick. There are 3000 linear arrays,  each containing 95 nanomagnets (total of 285,000 nanomagnets), and the ends of the nanostrips in each array are connected to two contact pads as shown in Fig. \ref{fig:SHNA}(a). Alternating charge current is passed between these two pads to generate alternating spin-orbit torques on the nanomagnets, which produce spin waves that radiate electromagnetic waves. The nanomagnets are made of Co, a magnetostrictive ferromagnet, and their thickness is 15 nm. Nanomagnet and ledge dimensions, intermagnet separation, etc. are all shown in the scanning electron micrograph in Fig. \ref{fig:SHNA}(b). The inter-nanomagnet separation is large enough that any dipole interaction between neighbors is negligible.

\subsection{Antenna activation}

An alternating current is pumped into each Pt nanostrip which injects spin currents of alternating spin polarization into the nanomagnets via the spin Hall effect \cite{dyakonov,hirsch} to cause alternating spin-orbit torque \cite{ralph} that results in either back-and-forth domain wall motion in the nanomagnets \cite{fert}, or magnetization precession, or both, which excites confined spin waves within the nanomagnets. The spin waves are time varying magnetizations (mimicking oscillating magnetic dipoles) that can radiate electromagnetic waves in the surrounding medium. The radiation  can be detected with receiving antennas.

There is a reason why the nanomagnets are ledged and the Pt strip is placed only over the ledges. Since Co is magnetostrictive, it physically expands and contracts when its magnetization alternates with the frequency of the pumped ac current. If we place the Pt strip directly on the nanomagnets to generate the alternating spin-orbit torque, it will ``clamp'' the nanomagnets and prevent the expansion/contraction, which will quench the spin waves and encumber the radiation process. It is imperative to {\it not} clamp the nanomagnets if they are magnetostrictive. This is the reason that the Pt strip is placed only on the ledges so as to not encumber the expansion/contraction of the bulk of the nanomagnets. The bottom surfaces of the nanomagnets are of course clamped by the underlying substrate, but this does not hinder the expansion/contraction of the top layers and hence does not prevent the generation of confined spin waves within the nanomagnets, which radiate electromagnetic waves.

 Of course, the ledges would have been entirely unnecessary if we had chosen a non-magnetostrictive or weakly magnetostrictive ferromagnet like Fe (instead of Co), which would not have experienced the expansion/contraction. We did not do that for a reason. If we can make the nanomagnets expand and contract with the frequency of the ac current (which requires sufficient magnetostriction), then this expansion/contraction will generate a time-varying strain in the substrate underneath, which will cause a surface acoustic wave (SAW) of the same frequency as the ac current to propagate in the substrate. That SAW can be picked up with interdigitated transducers if the substrate is piezoelectric (this is the reason for using the LiNbO$_3$ substrate; otherwise Si would work just as well). This will make this construct act as a {\it dual electromagnetic and acoustic antenna} that radiates an electromagnetic wave into space while simultaneously radiating an acoustic wave in the underlying substrate. The acoustic antenna functionality was already demonstrated by us in the past \cite{ahsanul} and hence not addressed here.

\subsection{Spin waves in the nanomagnets due to ac spin Hall effect caused by the ac current}

\begin{figure}[!h]
\centering
\includegraphics[width=0.9\textwidth]{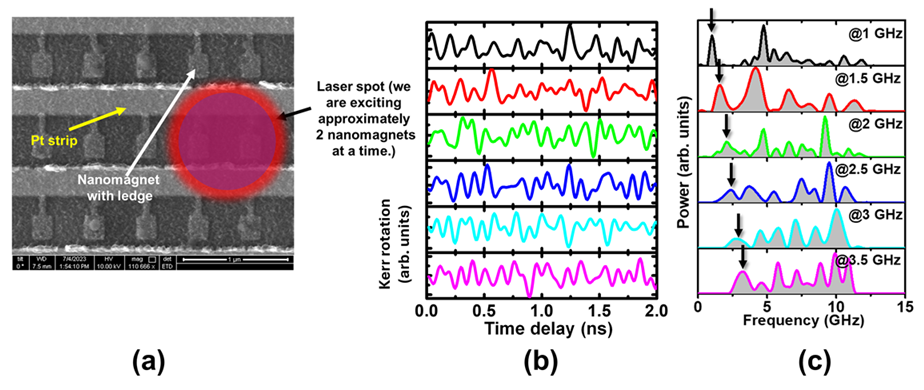}
\caption{(a) The pump and probe laser spots are approximately 1 $\mu$m in diameter and hence cover two nanomagnets at a time. Hence the spin waves are always sampled from two nanomagnets. (b) Kerr oscillations in the nanomagnets plotted in the time domain. (c) Fast Fourier transform of the Kerr oscillations showing the peaks in the spin wave spectra.}
\label{fig:Kerr}
\end{figure}

To further ascertain that pumping an alternating current into the heavy metal nanostrip overlying the ledges of the nanomagnets indeed excites spin waves (magnons) in them, we carried out time-resolved magneto-optical Kerr effect (TR-MOKE) measurements to confirm that spin waves are generated in the nanomagnets, and then find their frequency spectra. The details of the TR-MOKE setup are provided in the Supporting Information. We pumped alternating current into the heavy metal nanostrip at six different frequencies of 1, 1.5, 2, 2.5, 3 and 3.5 GHz with constant input power of 16 dbm. The pump and probe beam laser spots of the TR-MOKE setup overlap in space and cover only two nanomagnets at a time and hence we are sampling the spin waves from only two nanomagnets, as shown in Fig. \ref{fig:Kerr}(a). The measured Kerr-oscillations are shown in Fig. \ref{fig:Kerr}(b). Fast Fourier transform of these oscillations reveals the peaks in the spectra of the spin waves excited by the ac current, which are shown in Fig. \ref{fig:Kerr}(c). It is interesting to note that while there is always a peak at the current pumping frequency (as expected), there are also other peaks at higher frequencies which are not necessarily integral multiples of the pumping frequencies.  These other peaks are obviously not associated with the ac current injection since they occur at arbitrary frequencies. Further investigation showed that the frequencies where these peaks occur vary from one region of the nanomagnet array to another. Recall that the pump and probe beam of the TR-MOKE setup samples only two nanomagnets at a time; hence, we are always probing the modes locally, i.e. within those two nanomagnets. If we focus the pump and probe beam on a {\it different} nanomagnet pair, then the mode at the ac current frequency remains unchanged in frequency, but the other mode frequencies change. This is shown in the Supporting Information. Thus, the higher frequency modes are different in different regions of the nanomagnet array and ensemble averaging over the entire array will wash them out. Therefore, we do not expect to see radiation at their frequencies, but instead expect to see radiation only at the frequency of the ac current since ensemble averaging does not wash it out.

We cannot confirm the origin of the spatially-varying high frequency modes conclusively, but very likely they are vortex modes caused by strain pulses generated in the nanomagnet by the laser heating and cooling. The heating and cooling by the pump and probe pulses will cause strain pulses in the nanomagnets because of the unequal thermal expansion coefficients of the nanomagnets and the substrate \cite{yahagi,sucheta,sreya}. It has been shown that such strain pulses spawn vortex modes in magnetostrictive nanomagnets \cite{cui} and that the spectra of these modes depend on the nanomagnet diameter ($D$) and thickness ($t$). Since both $D$ and $t$ vary somewhat across the nanomagnet array, we expect to see variance in the frequencies of the high frequency modes and that is exactly what we see. Because of this variance, if we ensemble average the high frequency modes across the entire nanomagnet array of 285,000 nanomagnets that are present in our samples, then their amplitudes will become negligible compared to that of the one occurring at the frequency of the ac current. As a result, we cannot observe any electromagnetic radiation with the frequencies of the spatially-varying high-frequency modes, but can only observe electromagnetic radiation at the alternating current frequency.

\subsection{Electromagnetic radiation spectrum}

\begin{figure}[!t]
\centering
\includegraphics[width=0.99\textwidth]{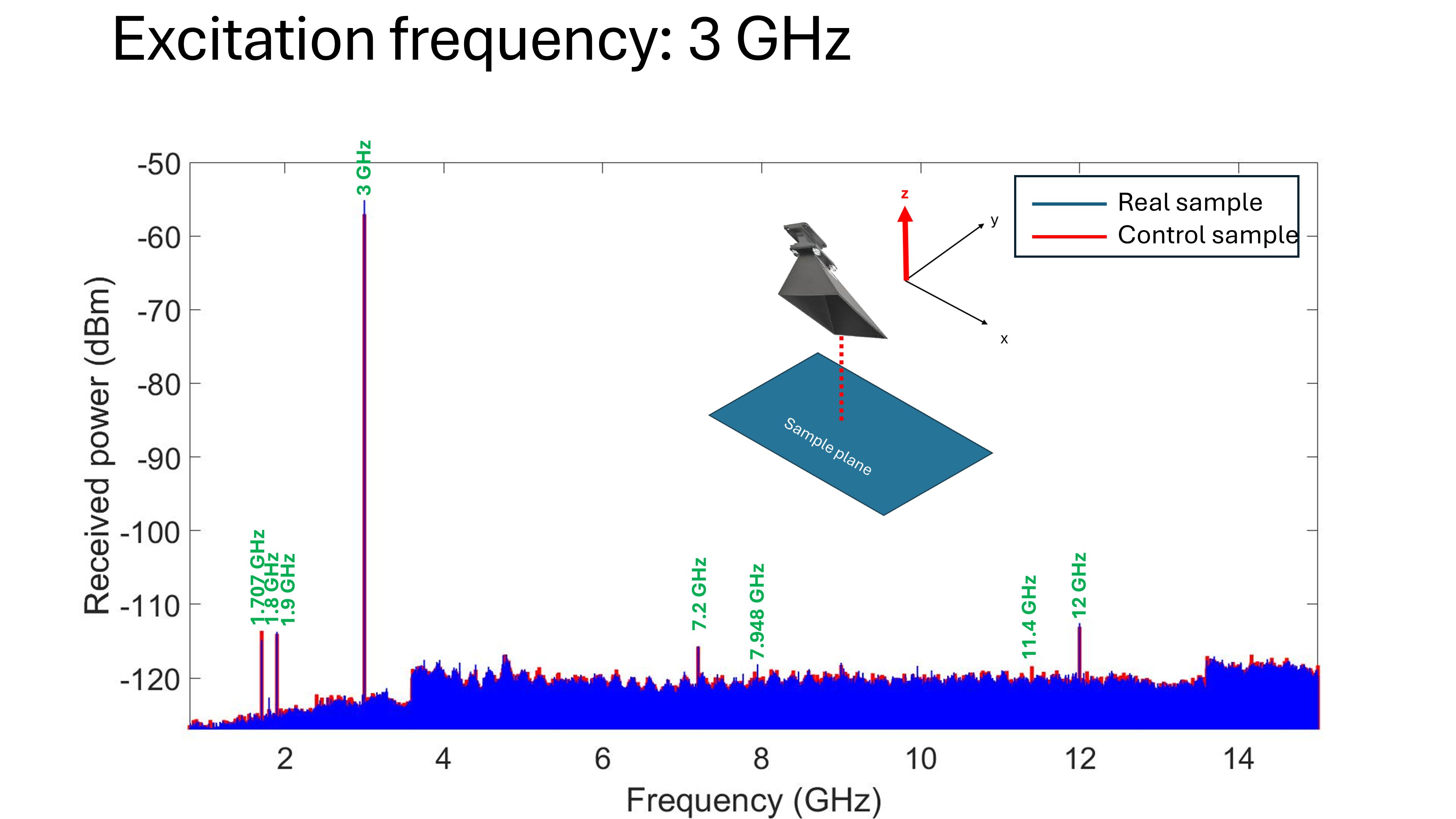}
\caption{Electromagnetic radiation spectrum of the real and control samples when the frequency of the ac current pumped into the Pt nanostrip is set to 3 GHz and the input power is 15 dbm. The distance between the receiving horn antenna and the sample in this case is 81 cm which is 8 times the wavelength, ensuring that we are measuring the far-field radiation.}
\label{fig:spectrum}
\end{figure}

We measured the spectra of the electromagnetic radiation emitted by the samples in an anechoic chamber, with the detecting horn antenna facing the plane of the nanomagnets as shown in the inset of Fig. \ref{fig:spectrum}. A spectrum analyzer was connected to the horn antenna to measure the spectrum of the received emission. The sample was placed at a distance of 81 cm from the horn antenna to ensure that we are always measuring the far-field radiation at the excitation frequency. The input power from the ac current source was set to 15 dbm (31 mW).

We had also measured the spectrum of the scattering parameter $S_{11}$ with a vector network analyzer and that data are shown in the Supporting Information. There are two sharp notches in the $S_{11}$ spectrum at 3.1 and 5.8 GHz, showing stronger coupling from the alternating current source into the sample at these frequencies. Because of this, we measured the spectrum of the emitted radiation with the driving alternating current frequency set to 3 GHz. 

The radiation measured by the horn antenna in the anechoic chamber is of course not solely due to the nanomagnets, but also has contributions from the Pt nanostrip, the contact lines, the contact pads and any other extraneous source of radiation. In an attempt to separate out the extraneous contributions, we fabricated two sets of samples that are otherwise nominally identical, except one has nanomagnets and the other does not. We call the latter ``control sample'' and measure the spectrum of the radiation it emits for comparison with that of the real sample which contains the nanomagnets. In Fig. \ref{fig:spectrum} we show the measured spectra from both the real and the control sample. Screenshots of the spectra taken from the spectrum analyzer can be found in the Supporting Information. 

It is important to understand that the difference between the power received at any point in space from the real and the control sample need not be positive because of {\it interference}. The wave emitted from the nanomagnets and that emitted by the peripherals interfere at any point in space.  Destructive interference will make the power received from the real sample (containing nanomagnets plus peripherals) {\it less} than that received from the control sample (containing only the peripherals), whereas constructive interference will have the opposite effect. This introduces an ambiguity that cannot be resolved easily. Furthermore, in our case, the observed difference is small in magnitude.
The power received from the real sample by the horn antenna of receiving area  3 cm $\times$ 3 cm was  -55 dbm (3.16 nW) whereas that received from the control sample was -57.5 dbm (1.77 nW). The difference of 1.4 nW, albeit well above the noise floor in the anechoic chamber [which was -120 dbm (1 pW)], is too small to allow us to make any quantitative estimates, given the limited sensitivities of the measuring equipment. We also cannot determine how much of this difference actually accrues from the nanomagnets and how much is due to unavoidable slight differences between the peripherals (i.e. contact pads, etc.) in the real sample and control sample.  Furthermore, as we show in the next section, the nanomagnets radiate anisotropically, so that the difference between the power received from the real sample and the control sample will be different in different directions for the same source-detector separation. All this prevents us from drawing any quantitative inferences and hinders us from determining the radiation efficiency of the SHNA.

We also notice that there are satellite peaks in the radiation spectrum that are present in both the real sample and the control sample. They are much weaker than the main peak. These satellite peaks do {\it not} conform to the higher frequency spin wave modes observed in Fig. \ref{fig:Kerr}. Furthermore, the control sample, which does not have any nanomagnets, also emits those peaks. This tells us that these satellite peaks are either from extraneous sources, or generated by the peripherals, and are {\it not} associated with the vortex modes generated in the nanomagnets by laser heating/cooling. As already mentioned, those vortex modes ensemble average out when averaged over the entire nanomagnet array and hence do not contribute to measurable radiation.

We conclude this section by pointing out that it is well-known in the context of conventional antennas that the gain, bandwidth and radiation efficiency will plummet if the antenna dimension is shrunk to small fractions of the electromagnetic wavelength that it radiates \cite{Harrington,skrivervik}. However, that happens only if the antenna is actuated by electromagnetic resonance. There is a school of thought that believes that if an antenna is actuated by acoustic resonance instead of electromagnetic resonance, then the effective wavelength that matters will be the acoustic wavelength at the frequency of radiation and not the electromagnetic wavelength. Since the former is typically five orders of magnitude smaller than the latter, this will allow antennas to be miniaturized to small fractions of the {\it electromagnetic} wavelength without sacrificing radiation efficiency or gain. This prompted significant research in acoustically actuated magneto-electric antennas of various types \cite{carman,northeastern,justine,raisa,saibal}. Their intrinsic radiation efficiencies did beat the theoretical limit on the radiation efficiencies of conventional antennas actuated by electromagnetic resonance, sometimes even by many orders of magnitude, which lends credence to this idea. The SHNA that is presented here, however, is {\it not} actuated acoustically and is of a completely different flavor. This new genre of antennas can open a new direction of
research in electrically small antennas whose dimensions are orders of magnitude smaller than the electromagnetic wavelength and yet they
radiate efficiently.

\section{Radiation patterns}

\begin{figure}[!hbt]
\centering
\includegraphics[width=0.9\textwidth]{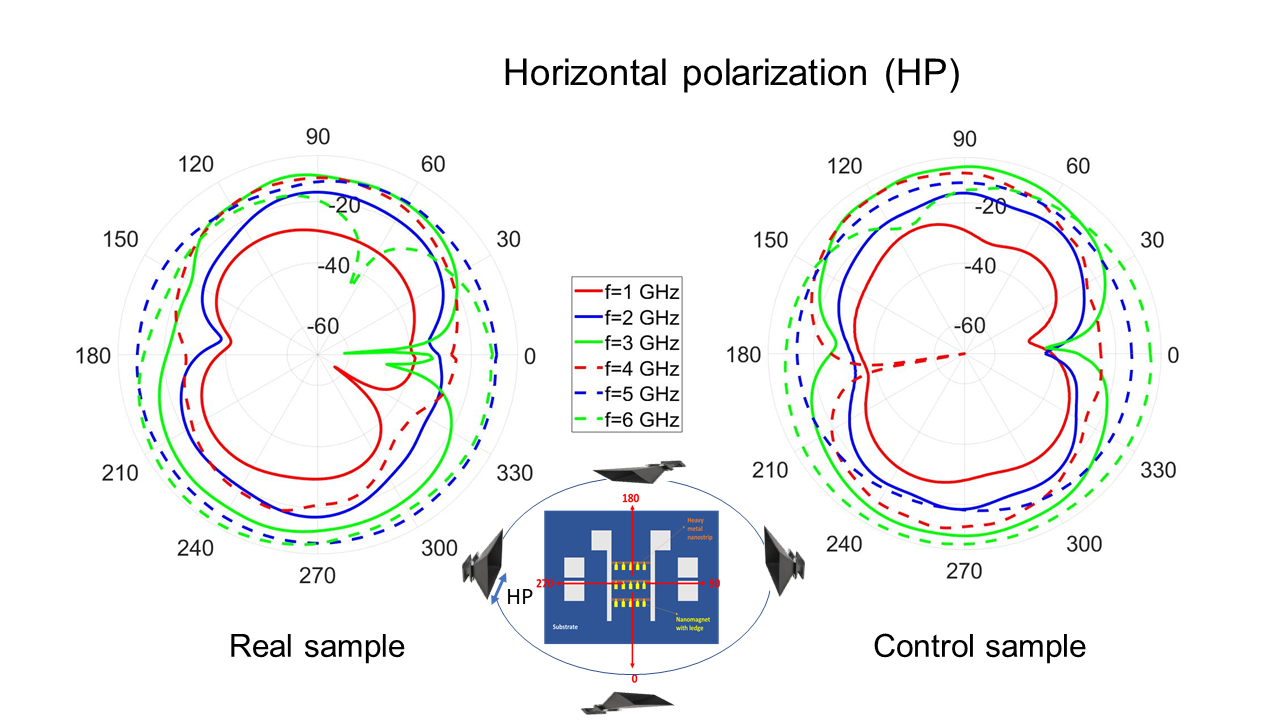}
\includegraphics[width=0.9\textwidth]{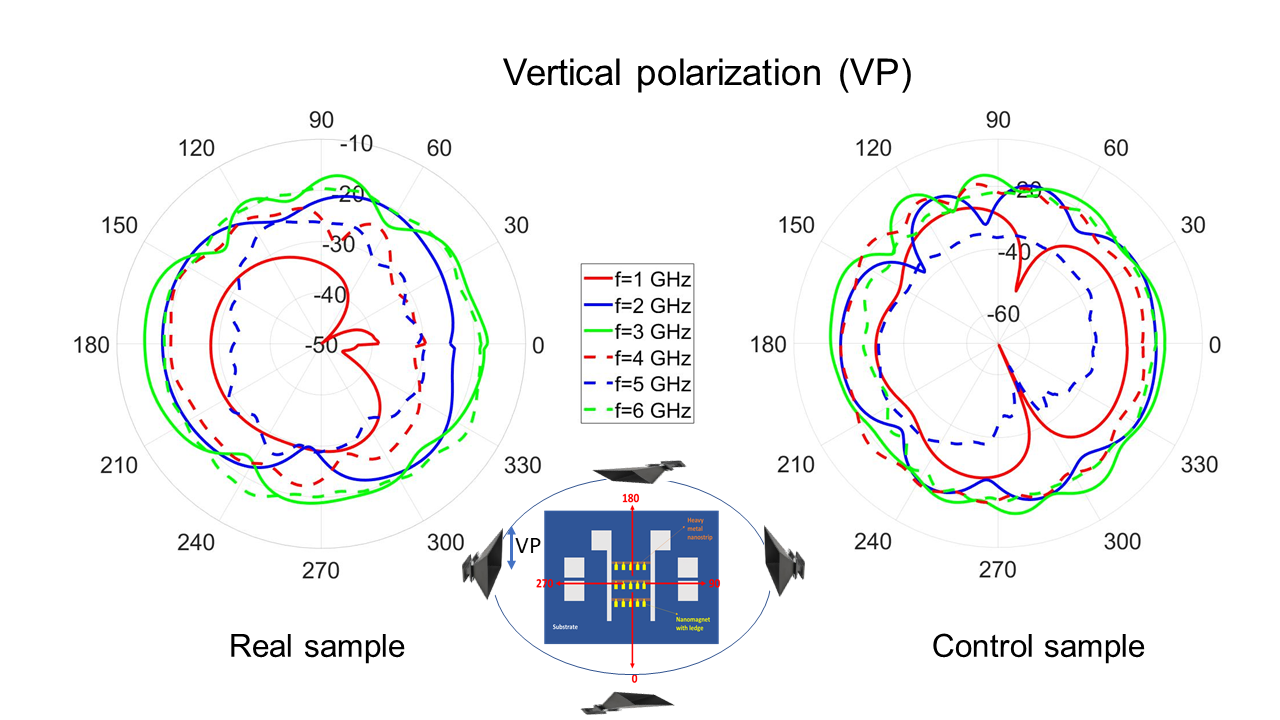}
\caption{The radiation pattern at different frequencies in the plane of the nanomagnets. The patterns are shown for both the real sample and the control sample, as well as for both horizontal and vertical polarizations.}
\label{fig:pattern}
\end{figure}

We measured the radiation patterns of the SHNA in three different planes -- the plane of the nanomagnets and the two transverse planes. They were measured at frequencies of 1, 2, 3, 4, 5 and 6 GHz (wavelengths 5 - 30 cm). The detector was always placed at a distance of 81 cm from the sample, which means that we are detecting the far-field radiation pattern. The patterns were measured for both the real sample and the control sample. The results are shown in Fig. \ref{fig:pattern} for both horizontal and vertical polarizations in the plane of the nanomagnets. The radiation patterns in the two transverse planes can be found in the Supporting Information.

Since the radiation pattern of the real sample is very significantly different from that of the control sample, we can once again conclude that the nanomagnets are radiating. This is further confirmatory evidence of the spin Hall nano-antenna (SHNA) operation. 

We note from Fig. \ref{fig:pattern} that the difference between the radiation from the real sample and the control sample varies quite strongly with direction, which means that the nanomagnets are radiating {\it anisotropically}. This anisotropy is surprising since the lateral dimension of the entire nanomagnet array ($\sim$ 160 $\mu$m) is much smaller than the electromagnetic wavelength at all measurement frequencies. Hence the entire nanomagnet array could be viewed as a point source that should radiate isotropically. Yet, it does not. To understand the origin of the anisotropy which is counter-intuitive, we have simulated the magnetization oscillations (spin waves) in the nanomagnets when pumped by the ac spin-orbit torque. 

\begin{figure}[!h]
\centering
\includegraphics[width=0.9\textwidth]{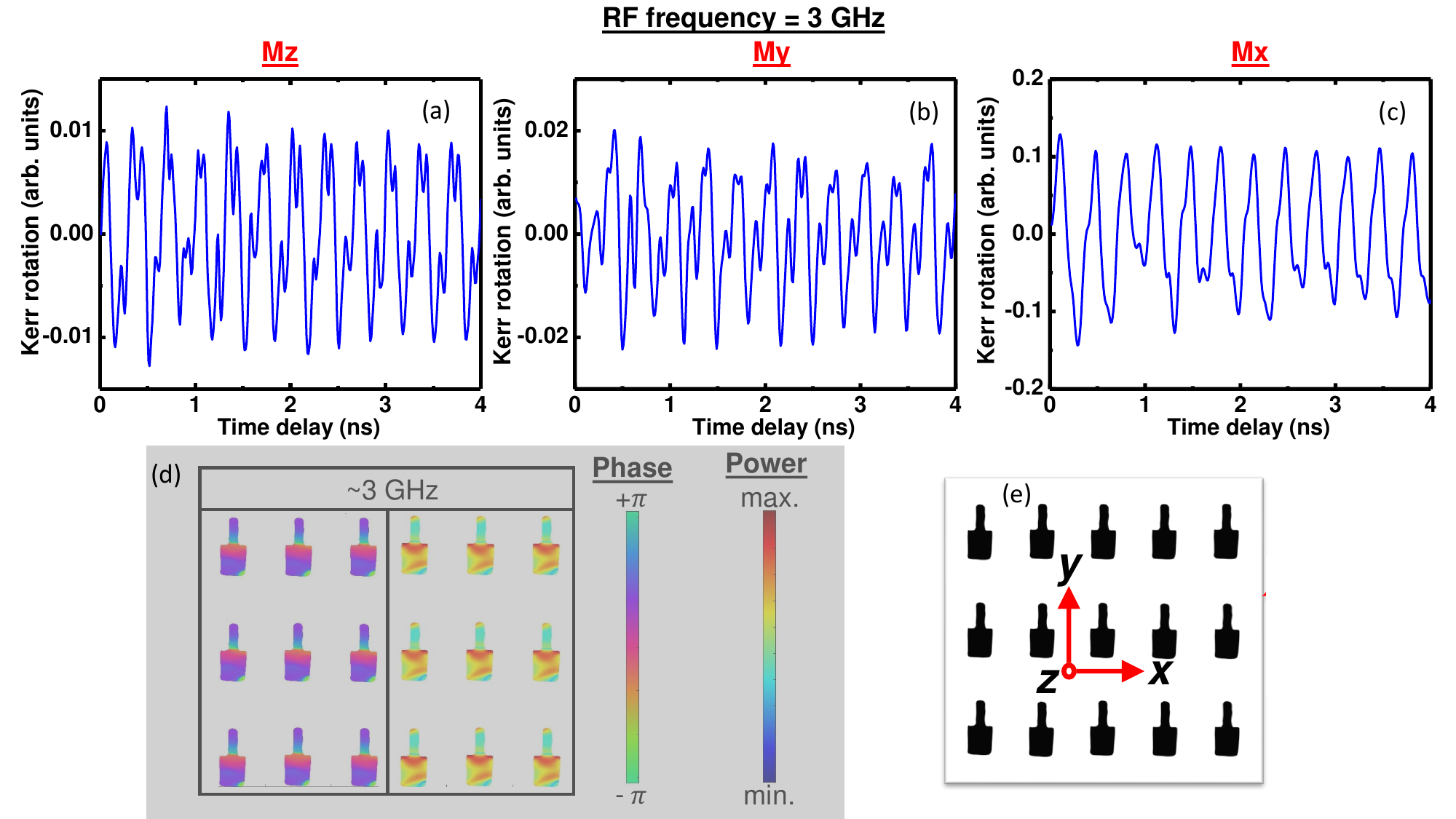}
\caption{Simulated magnetization oscillations (spin waves) in a nanomagnet at an ac current frequency of 3 GHz. (a) z-component (along the thickness); (b) y-component (along the ledge); (c) x-component (perpendicular to the ledge; (d) power and phase profiles of the spin waves inside a nanomagnet; (e) designation of the coordinate axes $x$, $y$ and $z$. }
\label{fig:oscillations}
\end{figure}

We used the micromagnetic simulator OOMMF package to find the magnetization components along the three coordinate axes $M_x(t)$, $M_y(t)$ and $M_z(t)$ as a function of time for three different frequencies of the pumping current: 3, 4 and 6 GHz. These plots are shown in Fig. \ref{fig:oscillations} for 3 GHz. It is interesting to note that the oscillations (and hence the spin waves) have much larger amplitude along the in-plane direction  that is perpendicular to the ledges (i.e., the $x$-direction). This feature is observed at all three frequencies of 3, 4 and 6 GHz. Only the 3 GHz results are shown in Fig. \ref{fig:oscillations}, while the other two can be found in the Supporting Information. Hence, the so-called ``point source'' has {\it internal anisotropy} since the spin waves have different amplitudes in different directions. That causes the anisotropy in the radiation pattern. Of course, this happens because of the anisotropic shape and geometry of the nanomagnets employed here. 

In Fig. \ref{fig:oscillations} we also show the power and phase profiles of the spin waves excited in the nanomagnet at 3 GHz excitation, calculated by the procedure described in \cite{barman}. Surprisingly, these profiles are quite frequency-dependent. For example, at 3 GHz, the spin wave power is concentrated at the edge of the nanomagnet abutting the ledge, while at 4 and 6 GHz (see Supporting Information) it is concentrated in the ledge and the edge facing away from the ledge. These features are responsible for the {\it frequency dependence} of the radiation pattern.

\section{Conclusion}
We have demonstrated a spin Hall nano-antenna (SHNA) which is an analogue of the celebrated spin Hall nano-oscillator (SHNO) -- the difference being that the latter delivers an alternating signal to a load whereas the former radiates an alternating signal into the surrounding medium. The SHNA is actuated by periodic spin-orbit torque (SOT) generated in nanomagnets by pumping an alternating current into a heavy metal nanostrip in physical contact with the nanomagnets. The alternating SOT, arising from the alternating spin Hall effect in the heavy metal, generates confined spin waves (magnons) in the nanomagnets that radiate electromagnetic waves (photons) in space.

 The SHNA can be orders of magnitude smaller than the radiated electromagnetic wavelength and yet emit efficiently. Hence, it is ideal for embedded applications where the antenna has to be much smaller than the wavelength (medically implanted devices, underwater/under-ice communication, wearable electronics, etc.) and yet must radiate with acceptable efficiency. We have also shown that while these antennas can be viewed as ``point sources'' because they are much smaller than the electromagnetic wavelength, they radiate anisotropically because the spin waves that are excited in the nanomagnets are anisotropic in nature.

\section*{Data Availability} 

All data generated are already available in the paper.

\section*{Acknowledgements}

R. F., S. B. and E. T. received support from the US National Science Foundation under grant ECCS-2235789. S. B. also received support from the Virginia Commonwealth University Commercialization Fund. A.B. gratefully acknowledges Department of Science and Technology, Govt. of India (grant no. DST/NM/TUE/QM-3/2019-1C-SNB) for financial assistance. A.B. and S.B. acknowledge support from the Indo-US Science and Technology Fund Center grant “Center for Nanomagnetics for Energy-Efficient Computing, Communications, and Data Storage” (IUSSTF/JC-030/2018).  P.K.P. acknowledges the Council of Scientific and Industrial Research (CSIR), Govt. of India for senior research fellowship.

\section*{Author contributions}
R. F. fabricated the samples. R. F. and M. S. made the antenna measurements. P. K. P 
 and A. K. M. carried out the TR-MOKE measurements. S. B., E. T. and A. B. supervised the project and verified the data. S. B.
conceived the idea. All authors contributed to writing the paper. R. F., M. S. and P. K. P. contributed equally.

\section*{Competing Interests}
The authors declare no competing interest.


\begin{thebibliography}{10}

\bibitem{dyakonov}
Dyakonov, M. I. \& Perel, V. I. Current-induced spin orientation of electrons in semiconductors. {\it Phys. Lett. A} {\bf 35}, 459 (1971).
\bibitem{hirsch}
Hirsch, J. E. Spin Hall effect. {\it Phys. Rev. Lett.} {\bf 83}, 12660 (1999).
\bibitem{chen}
Chen, T. et al. Spin torque and spin Hall nano-oscillators. {\it Proc. IEEE}  {\bf 104}, 1919-1945 (2016).
\bibitem{liu}
Liu, L., Pai, C. F., Li, Y., Tseng, H. W., Ralph, D. C. \& Buhrman, R. A. Spin-torque switching with the giant spin Hall effect of tantalum. {\it Science}, {\bf 336}, 555 (2012).
\bibitem{mellnik}
Mellnik, A. R., et al., Spin-transfer torque generated by a topological insulator. {\it Nature}, {\bf 511}, 449 (2014).
\bibitem{memory}
See, for example, Shao, Q., et al., Roadmap of spin-orbit torques. {\it IEEE Trans. Magn.}, {\bf 57}, 800439 (2021).
\bibitem{logic}
See, for example, Zhao, M. K., Wan, C. H., Luo, X. M., Wang, Y. Z., Ma, T. Y., Yang, W. L., Zhang, Y., Yin, L., Yu, G. Q. \& Han, X. F.
 Field-free programmable spin logics based on spin Hall effect. {\it Appl. Phys. Lett.}, {\bf 119}, 212405 (2021).
 \bibitem{MAMR}
 Okamoto, S., Kikuchi, N., Furuta, M.,  Kitakami, O. \& and Shimatsu, T. Microwave assisted magnetic recording technologies and related physics. {\it J. Phys. D: Appl. Phys.}, {\bf 48}, 353001 (2015).
\bibitem{abeed}
Abeed, M. A., Drobitch, J. L. \& Bandyopadhyay, S. Microwave oscillator based on a single straintronic magnetotunneling junction. {\it Phys. Rev. Appl.} {\bf 11}, 054069 (2019).
\bibitem{fert}
Khvalkovskiy, A. V., Cros, V., Apalkov, D., Nikitin, V., Krounbi, M., Zvezdin, K. A., Anane, A., Grollier, J \& Fert, A. Matching domain wall configuration and spin orbit torques for efficient domain wall motion. {\it Phys. Rev. B} {\bf 87}, 020402(R) (2013).
\bibitem{ahsanul}
Abeed M. A. \& Bandyopadhyay, S. Experimental Demonstration of an extreme subwavelength nanomagnetic acoustic antenna actuated by spin–orbit torque from a heavy metal nanostrip. {\it Adv. Mater. Technol.} {\bf 5}, 1901076 (2020).
\bibitem{litvinenko}
Litvinenko, A., Sethi, P., Murapaka, C., Jenkins, A., Cros, V., Bortolotti, P., Ferreira, R., Dieny, B. \& Ebels, U. Analog and digital phase modulation and signal transmission with spin-torque
nano-oscillators. {\it Phys. Rev. Appl.}, {\bf 16}, 024048 (2021).
\bibitem{manfrini1}
Manfrini, M., Devolder, T., Kim, J. V., Crozat, P.,  Zerounian, N., Chappert, C.,
van Roy, W., Lagae, L., Hrkac, G. \& Schrefl, T. Agility of vortex-based nanocontact spin torque oscillators. {\it Appl. Phys. Lett.}, {\bf 95}, 192507
(2009).
\bibitem{manfrini2}
Manfrini, M., Devolder, T., Kim, J. V., Crozat, P.,  Chappert, C., van Roy, W. \& Lagae, L. Frequency shift keying in vortex-based spin torque oscillators. {\it J. Appl. Phys.}, {\bf 109}, 083940 (2011).
\bibitem{pufall}
Pufall, M. R., Rippard, W. H., Kaka, S., Silva, T. J. \& Russek, S. E. Frequency modulation of spin-transfer oscillators. {\it Appl. Phys. Lett.}
{\bf 86}, 082506 (2005).
\bibitem{muduli}
Muduli, P. K., Pogoryelov, Y., Bonetti, S., Consolo, G., Mancoff, F. \& {\AA}kerman, J. Nonlinear frequency and amplitude modulation of a nanocontact-based spin-torque oscillator.
{\it Phys. Rev. B}, {\bf 81}, 140408 (2010).
\bibitem{pogoryelov}
Pogoryelov, Y., Muduli, P. K., Bonetti, S., Mancoff, F \& {\AA}kerman, J. Spin-torque oscillator linewidth narrowing under current modulation. {\it Appl. Phys.
Lett.}, {\bf 98}, 192506 (2011).
\bibitem{ralph}
Zhu, L., Ralph, D. C. \& and Buhrman, R. A. Maximizing spin-orbit torque generated by the spin Hall effect of Pt. {\it Appl. Phys. Rev.} {\bf 8}, 031308 (2021).
\bibitem{yahagi}
Yahagi, Y., Hartenek, B., Cabrini, S. \& Schmidt, H. Controlling nanomagnet magnetization dynamics via magnetoelastic coupling. {\it Phys. Rev. B}  {\bf 90}, 140405(R) (2014).
\bibitem{sucheta}
Mondal, S., Abeed, M. A., Dutta, K., De, A., Sahoo, S., Barman, A. \& Bandyopadhyay, S. Hybrid magnetodynamical modes in a single magnetostrictive nanomagnet on a piezoelectric substrate arising from magnetoelastic modulation of precessional dynamics. {\it ACS Appl. Mater. Interfaces} {\bf 10}, 43970 (2018).
\bibitem{sreya}
Pal, S., Pal, P. K., Fabiha, R., Bandyopadhyay, S. \& Barman, A. Acousto-Plasmo-Magnonics: Coupling spin waves with hybridized phonon-plasmon waves in a 2D artificial magnonic crystal deposited on a plasmonic material. {\it Adv. Funct. Mater.} {\bf 33}, 2304127 (2023).
\bibitem{cui}
Cui, H., Yang, X., Ni, L., Zhang, M., Liu, J., Wei, B., Chen Y. \& Yuan, J. High-frequency spin wave modes excited by strain pulse in vortex state magnetostrictive nanomagnets. {\it AIP Advances} {\bf 11}, 125314 (2021).
\bibitem{Harrington}
Harrington, R. F. Effect of antenna size on gain, bandwidth, and efficiency. {\it J. Res. Nat. Bur. Stand.} {\bf 64}, 1-12 (1960).
\bibitem{skrivervik}
Skrivervik, A. K.,  Z\"urcher, J. F., Staub, O., J. \& Mosig, J. R. PCS antenna design: The challenge of miniaturization. {\it IEEE Antennas
Propag. Mag.} {\bf 43}, 12 (2001).
\bibitem{carman}
Domann J. P. \& Carman, G. P. Strain powered antennas. {\it J. Appl. Phys.} {\bf 121}, 044905 (2017) and references therein.
\bibitem{northeastern}
Nan, T.,  et al. Acoustically actuated ultra-compact NEMS magnetoelectric antennas.  {\it Nat. Commun.}, {\bf 8}, 296 (2017) and references therein.
\bibitem{justine}
Drobitch, J. L., De, A., Dutta, K., Pal, P. K., Adhikari, A., Barman, A. \& Bandyopadhyay, S. Extreme subwavelength magnetoelastic electromagnetic antenna implemented with multiferroic nanomagnets. {\it Adv. Mater. Technol.} {\bf 5}, 2000316 (2020).
\bibitem{raisa}
Fabiha, R., Lundquist, J., Majumder, S., Topsakal, E., Barman, A. \& Bandyopadhyay, S. Spin wave electromagnetic nano-antenna enabled by tripartite phonon-magnon-photon coupling. {\it Adv. Sci.} {\bf 9}, 2104644 (2022).
\bibitem{saibal}
Samanta, A. \& Roy, S. Generation of microwaves With tunable frequencies in ultracompact “Magnon Microwave Antenna” via phonon-magnon-photon coupling.
{\it IEEE Trans. Elec. Dev.} {\bf 70}, 335 (2022).
\bibitem{barman}
Kumar, D., Dmytriiev, O., Ponraj, S. \& Barman, A. Numerical calculation of spin wave dispersions in magnetic nanostructures. {\it J. Phys. D: Appl. Phys} {\bf 45}, 015001 (2012).
\end{thebibliography}
\end{document}



\title{(Supplemental Material)\\
Tripartite Phonon-Magnon-Plasmon Coupling, Parametric Amplification, and Formation of a Phonon-Magnon-Plasmon Polariton in a Two-Dimensional Periodic Array of Magnetostrictive/Plasmonic Bilayered Nanodots}
}

\author{Sreya Pal}
\affiliation{Department of Condensed Matter and Materials Physics, S. N. Bose National Centre for Basic Sciences, Block–JD, Sector–III, Salt Lake, Kolkata, 700106, India}

\author{Pratap Kumar Pal}
\affiliation{Department of Condensed Matter and Materials Physics, S. N. Bose National Centre for Basic Sciences, Block–JD, Sector–III, Salt Lake, Kolkata, 700106, India}

\author{Raisa Fabiha}
\affiliation{Department of Electrical and Computer Engineering,
Virginia Commonwealth University, Richmond, VA 23284, USA}

\author{Supriyo Bandyopadhyay}%
\email{sbandy@vcu.edu}
\affiliation{Department of Electrical and Computer Engineering,
Virginia Commonwealth University, Richmond, VA 23284, USA}

\author{Anjan Barman}%
 \email{abarman@bose.res.in}
\affiliation{Department of Condensed Matter and Materials Physics, S. N. Bose National Centre for Basic Sciences, Block–JD, Sector–III, Salt Lake, Kolkata, 700106, India}

\maketitle

\textcolor{black}{\section{Position dependence of the reflectivity spectra for plasmonic and non-plasmonic samples}
To ascertain that the observed differences between the
reflectivity spectra of the Al-containing and Al-lacking
samples are indeed intrinsic (i.e., due to the presence and
absence of plasmons), and not a mere artifact of extrinsic
sample differences such as unintentional variations in
size, shape, material composition, etc., we have measured
the reflectivity oscillations and the corresponding spectra
in four different regions in each wafer (one with Al
and the other without) by focusing the TR-MOKE laser
spot in those four different regions as shown in Fig.S1.}
\begin{figure*}[htbp!]
\centering
\includegraphics[width=13cm]{S1.jpg}
\textcolor{black}{\caption{(a) Low resolution scanning electron microscope (SEM) images of a wafer showing an array of squares each containing
nanomagnets. (b) A higher resolution image of a randomly picked square showing the laser focused in four different regions. The laser spot covers approximately 36 nanomagnets. The black regions are alignment marks for navigation of the nanodot array across the wafer.}}
\label{FigS1_re}
\end{figure*}

\textcolor{black}{The results are presented in Fig. S2 and Fig. S3 for the Al-lacking and Al-containing samples respectively. The Al-lacking samples show the same behavior in all four probed regions (always 2 peaks in the spectra occurring at about the same frequencies) as do the Al-containing samples (always 4 peaks, again occurring at about the same frequencies). Their difference is remarkably consistent. This confirms that the observed differences in the spectra between the plasmonic and non-plasmonic specimens are intrinsic and not extrinsic.}

\begin{figure*}[htbp!]
\centering
\includegraphics[width=13cm]{S2.jpg}
\textcolor{black}{\caption{(a) Background-subtracted time-resolved data for the reflectivity and (b) the corresponding FFT in four “samples” devoid of Al. The four different “positions” refer to four different regions of the wafer, or effectively four different “samples” each consisting of $\sim$36 nanomagnets. The spectra are remarkably similar in all samples. There are always two peaks $(f\textsubscript{1})$ and $(f\textsubscript{2})$ that occur at about the same frequencies.}}
\label{FigS3}
\end{figure*}

\begin{figure*}[htbp!]
\centering
\includegraphics[width=12cm]{S3.jpg}
\textcolor{black}{\caption{(a) Reflectivity oscillations and (b) the corresponding FFT in four 'samples' containing Al. The four different 'positions' refer to four
different regions of the wafer, or effectively four different 'samples' each consisting of $\sim$36 nanomagnets. The spectra are remarkably similar in all samples. There are always four peaks $f\textsubscript{1}^{'}$, $f\textsubscript{2}^{'}$, $f\textsubscript{3}^{'}$ and $f\textsubscript{4}^{'}$that occur at about the same frequencies.}}
\label{FigS4}
\end{figure*}

\newpage
\section {Spin-wave mode analysis for Co nanomagnets (Non-plasmonic samples}
To understand the characteristics of three magnetic modes in the samples without Al, including the Kittel-type and other modes, we used an in-house simulator Dotmag \cite{Kumar12} to calculate their spatial properties (power and phase). The results, presented in Fig. \ref{FigS1}, reveal distinct features.
\begin{figure*}[htbp!]
\centering
\includegraphics[width=13cm]{S4.jpg}
\caption{Simulated power and phase profiles of different spin-wave modes (mode 1, mode 2, mode 3) for Co nanomagnets without Al at four bias magnetic fields. Colour maps for power and phase distributions are shown below the images.}
\label{FigS1}
\end{figure*}
For mode '2', the power profile indicates a uniform distribution of spin-wave (SW) power at the nanomagnet's center, resembling a uniform precessional mode or center mode (CM). Despite high power at the edges, the central distribution is significant. In contrast, mode '1' shows power concentrated at the edges, termed the edge mode (EM), with power focused at two vertical edges perpendicular to the bias field. Mode '3' is akin to '2', except the quantization axis rotates about 45° with respect to the major axis of the elliptical nanomagnet. Analyzing phase profiles, we observe spins precessing alternately in opposite phases, creating azimuthal contrast with an azimuthal mode quantization number of 2. Notably, power distributions for these modes (both at the center and vertical edges of the nanomagnet) remain remarkably stable, showing minimal change with variations in external bias field strength.


\bibliography{Reference}